\documentclass[12pt]{article}
\usepackage{amsmath}
\usepackage{cite}
\usepackage{graphicx}
\usepackage{amsfonts}
\usepackage{amssymb}
\usepackage{appendix}
\providecommand{\U}[1]{\protect\rule{.1in}{.1in}}

\csname @addtoreset\endcsname{equation}{section}
\newcommand{\eq}{\begin{equation}}
\newcommand{\feq}{\end{equation}}
\newcommand{\eqn}{\begin{eqnarray}}
\newcommand{\feqn}{\end{eqnarray}}
\newcommand{\arr}{\begin{eqnarray*}}
\newcommand{\farr}{\end{eqnarray*}}

\newcommand{\bea}{\begin{eqnarray}}
\newcommand{\eea}{\end{eqnarray}}

\begin{document}
\begin{titlepage}
\begin{center}
\renewcommand{\thefootnote}{\fnsymbol{footnote}}
{\Large{\bf Metrics depending on one variable in D-dimensional Einstein-Maxwell Theory}}
\vskip1cm
\vskip 1.3cm
W. A. Sabra
\vskip 1cm
{\small{\it
Department of Physics, American University of Beirut, Lebanon }}
\end{center}
\bigskip
\begin{center}
{\bf Abstract}
\end{center}
We present new families of solutions of $D$-dimensional Einstein-Maxwell theory depending on one variable
for all space-time signatures. The solutions found can be thought of as generalized Melvin solutions including 
fluxtubes, domain walls and cosmological space-times. Explicit examples are given in four and five space-time dimensions.
\end{titlepage}
\section{Introduction}
Some years ago Kasner presented a family of Euclidean four-dimensional
vacuum solutions depending on one variable \cite{kasner}. The original form
of the metric, when generalized to all space-time signatures \cite{harvey},
is of the form 
\begin{equation}
ds^{2}=\epsilon_{0}x_{1}^{2a_{1}}dx_{1}^{2}+
\epsilon_{1}x_{1}^{2a_{2}}dx_{2}^{2}+\epsilon_{2}x_{1}^{2a_{3}}dx_{3}^{2}+
\epsilon_{3}x_{1}^{2a_{4}}dx_{4}^{2}
\end{equation}
with $\epsilon_{i}=\pm1.$ These are vacuum solutions of Einstein gravity
provided that the constants characterizing the solutions satisfy the
conditions
\begin{align*}
a_{2}+a_{3}+a_{4} & =a_{1}+1\text{ }, \\
a_{2}^{2}+a_{3}^{2}+a_{4}^{2} & =\left( a_{1}+1\right) ^{2}\text{ }.
\end{align*}
With a coordinate redefinition (or effectively setting $a_{1}=0),$ we obtain
the metric 
\begin{equation}
ds^{2}=\epsilon_{0}d\tau^{2}+\epsilon_{1}\tau^{2p_{1}}dx^{2}+\epsilon_{2}
\tau^{2p_{2}}dy^{2}+\epsilon_{3}\tau^{2p_{3}}dz^{2}  \label{fk}
\end{equation}
with the conditions on the so called Kasner exponents given by
\begin{equation}
p_{1}+p_{2}+p_{3}=p_{1}^{2}+p_{2}^{2}+p_{3}^{2}=1\text{ }.
\end{equation}
The solution with $\epsilon_{0}=-1$ and $\epsilon_{1}=\epsilon_{2}=
\epsilon_{3}=1$ is the Bianchi I metric \cite{macel} which is normally
referred to in the literature as the Kasner metric. Earlier solutions
related to Kasner metric are due to Weyl, Levi-Civita and Wilson. In fact,
the Kasner metric was re-discovered by many authors in various forms. For a
presentation and a discussion on all related solutions that appeared before
and after Kasner work and references, we refer the reader to \cite{harvey}.
We note that the flat metric corresponds to the case were (\ref{fk}) has two
vanishing exponents.

The nature of the solution depends on the choice of the parameters $
\epsilon_{i}$. The solutions (\ref{fk}) can describe static anisotropic
domain wall solutions or cosmological solutions. In the domain wall cases, $
\tau$ is taken to be the radial coordinate. More precisely, we take $
\epsilon_{0}=-\epsilon_{1}=\epsilon_{2}=\epsilon_{3}=1,$ and after
relabelling the coordinates, we obtain the anisotropic domain wall solutions 
\begin{equation}
ds^{2}=-r^{2p_{1}}dt^{2}+dr^{2}+r^{2p_{2}}dx^{2}+r^{2p_{3}}dy^{2}.
\label{gcy}
\end{equation}
The cosmological solutions with Lorentzian signature are simply given by the
Bianchi I metric 
\begin{equation}
ds^{2}=-dt^{2}+t^{2p_{1}}dx^{2}+t^{2p_{2}}dy^{2}+t^{2p_{3}}dz^{2}.
\end{equation}

It is of interest to find generalizations of Kasner metric for theories with
non-vanishing gauge fields. Kasner-like solutions, with fixed Kasner
exponents, were obtained in \cite{phantom} for four-dimensional
Einstein-Maxwell theory with a non-canonical sign for the Maxwell field
kinetic term.

A physically important solution of Einstein-Maxwell theory, depending on one
variable, with many applications in black hole physics and cosmology is the
Melvin fluxtube \cite{melvin}. In \cite{kt}, Kasner-like generalizations of
Melvin solution were constructed together with a study of a domain
wall/cosmology correspondence for spacetimes with a non-vanishing Maxwell
field. The domain walls constructed are generalizations of Melvin
space-times and their related magnetic cosmologies can be thought of as
generalizations of the Rosen solutions \cite{rosen}.

In \cite{s1} time-dependent and static solutions, in arbitrary space-time
dimensions and signatures, generalizing Kasner solutions to include matter
fields were considered. More specifically, solutions were found for all
known supergravity theories containing form fields and a coupled dilaton 
\cite{Hull}. Such results were later extended to $N=2$ supergravity theories
coupled to vector multiplets in four and five dimensions \cite{s2, s3}.

Our current work will be devoted to finding general exact solutions
depending on one variable in $D$-dimensional Einstein-Maxwell theory with
arbitrary space-time signatures. Like generalized Kasner metrics and Melvin
solutions, our new solutions will be of importance in the analysis of
cosmological singularity, gravitational turbulence, chaos \cite{book} as
well as black hole physics and cosmology.

In addition to Kasner-like metrics constructed in \cite{kt, s1}, we obtain
more general solutions depending on one variable. The new solutions are
determined in terms of restricted matrices. The number of the families of
solutions will depend, for a given space-time dimension, on all possible
canonical Jordan forms of these matrices.

We organize our work as follows. In the next section we analyze the
equations of motion of arbitrary signature $D$-dimensional Einstein-Maxwell
theory and construct two classes of solutions corresponding to two different
choices of the gauge field. Section three includes some explicit examples
and a summary.

\section{Einstein-Maxwell Solutions}
Our starting point is the $D$-dimensional gravity with a Maxwell field
described by the action 
\begin{equation}
S=\int d^{D}x\sqrt{\left\vert g\right\vert }\left( R-\frac{\varepsilon}{4}%
\,\,\mathcal{F}_{\mu\nu}\mathcal{F}^{\mu\nu}\right) \text{ },  \label{gact}
\end{equation}
with $\varepsilon=\pm1.$ The gravitational equations of motion derived from (
\ref{gact}) are given by 
\begin{equation}
R_{\mu\nu}-\varepsilon\left( \frac{1}{2}\mathcal{F}_{\mu\alpha}\mathcal{F}
_{\nu}{}^{\alpha}-g_{\mu\nu}\frac{1}{4(D-2)}\mathcal{F}_{\mu\nu }\mathcal{F}
^{\mu\nu}\,\right) =0  \label{eqm}
\end{equation}
and the Maxwell equation is given by%
\begin{equation}
\partial_{\mu}\left( \sqrt{\left\vert g\right\vert }\,\,\mathcal{F}^{\mu\nu
}\right) =0\text{ }.  \label{max}
\end{equation}
As already stated, we are interested in finding exact solutions depending on
one variable with a non-trivial Maxwell field. We start with the following
generic metric solution 
\begin{equation}
ds^{2}=\epsilon_{0}d\tau^{2}+g_{zz}(\tau)dz^{2}+g_{ab}(\tau)dx^{a}dx^{b}
\text{ }  \label{lan1}
\end{equation}
with $g_{za}=0$ and $a=1,...,D-2.$ First we consider solutions with only
non-vanishing $F^{\tau z}$ given by
\begin{equation}
\mathcal{F}^{\tau z}=\frac{Q}{\omega}\text{ },  \label{msol}
\end{equation}
where $Q$ is a constant and $\omega=\sqrt{\left\vert g\right\vert }$. The
analysis of the Einstein equations of motion (\ref{eqm}) gives
\begin{align}
\frac{\ddot{\omega}}{\omega}-\frac{\dot{\omega}^{2}}{\omega^{2}}+\frac{1}{4}
\kappa_{\text{ }j}^{i}\kappa_{\text{ }i}^{j}+\varepsilon\frac{\left(
D-3\right) }{2\left( D-2\right) }\frac{Q^{2}}{\omega^{2}}\,g_{zz} & =0\text{ 
},  \label{m1} \\
\dot{\kappa}^{z}{}_{z}+\kappa_{\text{ }z}^{z}\frac{\dot{\omega}}{\omega }
+\varepsilon\frac{\left( D-3\right) }{\left( D-2\right) }\frac{Q^{2}}{%
\omega^{2}}g_{zz} & =0\text{ },  \label{m2} \\
\dot{\kappa}^{a}{}_{b}+\kappa_{\text{ }b}^{a}\frac{\dot{\omega}}{\omega }
-\varepsilon\delta_{\text{ }b}^{a}\frac{1}{(D-2)}\frac{Q^{2}}{\omega^{2}}
g_{zz} & =0\text{ },  \label{m4}
\end{align}
where $\kappa_{\text{ }j}^{i}=g^{ik}\dot{g}_{kj}$\ \cite{landau} and the
overdot denotes the derivative with respect to the variable $\tau.$
Furthermore, equations (\ref{m2}) and (\ref{m4}) imply the relation 
\begin{equation}
\ddot{\omega}=\frac{\varepsilon Q^{2}}{2(D-2)\omega}g_{zz}\text{ }.
\label{tr}
\end{equation}
Using (\ref{tr}), then (\ref{m2}) and (\ref{m4}) can be integrated and we
obtain 
\begin{align}
\kappa^{z}{}_{z} & =\frac{1}{\omega}\left( \theta^{z}{}_{z}-\left(
2D-6\right) \dot{\omega}\right) \text{ },  \label{k1} \\
\kappa^{a}{}_{b} & =\frac{1}{\omega}\left( \theta_{\text{ }b}^{a}+2\delta_{
\text{ }b}^{a}\dot{\omega}\right) \text{ },  \label{k2}
\end{align}
where $\theta^{i}{}_{j}=\left( \theta^{z}{}_{z},\text{ }\theta_{\text{ }
b}^{a}\right) $ are constants satisfying the condition
\begin{equation}
\theta_{\text{ }i}^{i}=\theta_{\text{ }a}^{a}+\theta_{\text{ }z}^{z}=0\text{ 
}.  \label{conzero}
\end{equation}
Noting that $\kappa_{\text{ }j}^{i}=g^{ik}\dot{g}_{kj},$ we then obtain,
using (\ref{k1}) and (\ref{k2}), the following equations for the space-time
metric components
\begin{align}
\dot{g}_{zz} & =\frac{1}{\omega}\left( \theta^{z}{}_{z}-2\left( D-3\right) 
\dot{\omega}\right) g_{zz}\text{ },  \label{met1} \\
\dot{g}_{ab} & =\frac{1}{\omega}\left( g_{ac}\theta_{\text{ \ }b}^{c}+2\dot{
\omega}g_{ab}\right) \text{ }.  \label{met2}
\end{align}
Substituting (\ref{k1}) and (\ref{k2}) into (\ref{m1}), we obtain the
condition
\begin{equation}
2\left( D-2\right) \left( \theta ^{z}{}_{z}\dot{\omega}-\left( D-3\right) 
\dot{\omega}^{2}\right) -\varepsilon g_{zz}Q^{2}=\frac{1}{2}\theta _{\text{ }
j}^{i}\theta _{\text{ }i}^{j}\text{ }.  \label{ni}
\end{equation}
In summary, our metric solutions are entirely defined in terms of the
equations (\ref{tr}), (\ref{conzero}), (\ref{met1}, (\ref{met2}) and (\ref
{ni}). In what follows, we solve these equations by making the following
convenient change of variables 
\begin{equation}
\frac{d\sigma }{d\tau }=\frac{1}{H(\sigma )}\text{ },\text{ \ \ \ \ \ \ \ \ }
\omega =\sigma H(\sigma )\text{ }.  \label{cv}
\end{equation}%
The equations defining the space-time metric components (\ref{met1}) and ( 
\ref{met2}) then become 
\begin{eqnarray}
\frac{dg_{zz}}{d\sigma } &=&\left[ \frac{2}{\sigma }p-2(D-3)\frac{1}{H}\frac{
dH}{d\sigma }\right] g_{zz}\text{ }, \\
\frac{dg_{ab}}{d\sigma } &=&\left( \frac{2}{\sigma }g_{ac}\lambda _{\text{ \ 
}b}^{c}+\frac{2}{H}g_{ab}\frac{dH}{d\sigma }\right) \text{ },
\end{eqnarray}
where we have defined 
\begin{equation}
\lambda _{\text{ \ }z}^{z}=p=\frac{1}{2}\left( \theta _{\text{ }
z}^{z}+6-2D\right) \text{ },\text{ \ \ \ \ }\lambda _{\text{ }b}^{a}=\frac{1%
}{2}\left( \theta _{\text{ }b}^{a}+2\delta _{\text{ }b}^{a}\right) \text{ }.
\label{nc}
\end{equation}
These equations admit the solutions
\begin{align}
g_{zz} & =\epsilon_{zz}\sigma^{2p}H^{-2(D-3)}\text{ }, \\
g_{ab} & =h_{ac}\left( e^{2\lambda\log\sigma}\right) ^{c}{}_{b}H^{2}\text{ }.
\end{align}
with $\epsilon_{zz}=\pm1$ and the constants $h_{ac}$ are symmetric and
satisfying 
\begin{equation}
h_{ac}\lambda^{c}{}_{b}=h_{bc}\lambda^{c}{}_{a}\text{ }.  \label{sym}
\end{equation}
The function $H$ can be determined using (\ref{tr}), which in terms of the
new variables reads 
\begin{equation}
\frac{1}{H}\frac{dH}{d\sigma}-\frac{\sigma}{H^{2}}\left( \frac{dH}{d\sigma }
\right) ^{2}+\frac{\sigma}{H}\frac{d^{2}H}{d\sigma^{2}}=\frac{\varepsilon
\epsilon_{zz}Q^{2}}{2(D-2)}\sigma^{\left( 2p-1\right) }H^{-2(D-3)}\text{ }.
\end{equation}
This admits the solution 
\begin{equation}
H=\left( 1+\mu\sigma^{2p}\right) ^{\frac{1}{D-3}}
\end{equation}
with the condition
\begin{equation}
\mu=\varepsilon\epsilon_{zz}\frac{(D-3)}{8(D-2)}\left( \frac{Q}{p}\right)
^{2}\text{ }.  \label{z1}
\end{equation}
Plugging our solution back into the relation (\ref{ni}), we obtain
\begin{equation}
\frac{1}{2}\theta _{\text{ }j}^{i}\theta _{\text{ }i}^{j}=2(D-2)\left[
p-(D-3)\right] \text{ }.  \label{con2}
\end{equation}%
Using (\ref{nc}), the conditions (\ref{conzero}) and (\ref{con2}) become 
\begin{align}
\lambda _{\text{ }i}^{i}& =p+\lambda _{\text{ }a}^{a}=1\text{ },\text{ \ \ \
\ }  \label{z2} \\
\lambda _{\text{ }j}^{i}\lambda _{\text{ }i}^{j}& =p^{2}+\lambda _{\text{ }
b}^{a}\lambda _{\text{ }a}^{b}=1\text{ }.  \label{z3}
\end{align}
To summarize, our solutions are given by 
\begin{align}
ds^{2}& =\epsilon _{0}H^{2}d\sigma ^{2}+\epsilon _{zz}\sigma
^{2p}H^{-2(D-3)}dz^{2}+h_{ac}\left( e^{2\lambda \log \sigma }\right)
^{c}{}_{b}H^{2}dx^{a}dx^{b}  \notag \\
H& =\left( 1+\varepsilon \epsilon _{zz}\frac{(D-3)}{8(D-2)}\left( \frac{Q}{p}
\right) ^{2}\text{ }\sigma ^{2p_{\text{ }}}\right) ^{\frac{1}{D-3}}\text{ },
\text{ \ \ }\mathcal{F}^{\sigma z}=\frac{Q}{\sigma H^{2}}\text{ },
\label{gs}
\end{align}
supplemented with the conditions (\ref{sym}), (\ref{z2}) and (\ref{z3}).

In what follows, we consider a second class of solutions depending on one
variable with the $D$-dimensional metric%
\begin{equation}
ds^{2}=\epsilon_{0}d\tau^{2}+g_{zz}(\tau)dz^{2}+g_{ww}(\tau)d%
\omega^{2}+g_{ab}(\tau)dx^{a}dx^{b}
\end{equation}
with $g_{az}=g_{aw}=g_{zw}=0$. The only non-vanishing gauge field-strength
component for this class of solutions is given by
\begin{equation}
\mathcal{F}_{zw}=P  \label{a}
\end{equation}%
with constant $P$. A similar analysis to that of the previous case, gives
the following equations for the metric components 
\begin{align}
\dot{g}_{zz}& =\frac{1}{\omega }\left( \theta ^{z}{}_{z}+2\dot{\omega}
\right) g_{zz}\text{ },  \label{d1} \\
\dot{g}_{ww}& =\frac{1}{\omega }\left( \theta ^{w}{}_{w}+2\dot{\omega}
\right) g_{ww}\text{ },  \label{d3} \\
\dot{g}_{ab}& =\frac{1}{\omega }\left( g_{ac}\theta _{\text{ }b}^{c}-\frac{2
}{D-3}g_{ab}\dot{\omega}\right) \text{ },  \label{d2}
\end{align}
with 
\begin{equation}
\theta ^{z}{}_{z}+\theta ^{w}{}_{w}+\theta _{\text{ }a}^{a}=0  \label{w}
\end{equation}
and 
\begin{align}
\ddot{\omega}+\frac{\left( D-3\right) }{2\left( D-2\right) }\varepsilon
\epsilon _{0}P^{2}g^{zz}g^{ww}\omega & =0\text{ },  \label{tr3} \\
\varepsilon \epsilon _{0}g^{zz}g^{ww}P^{2}\omega ^{2}-2\left( \frac{D-2}{D-3}
\right) \left( \dot{\omega}^{2}-\theta ^{a}{}_{a}\dot{\omega}\right) -\frac{1
}{2}\theta ^{i}{}_{j}\theta ^{j}{}_{i}& =0\text{ }.  \label{in}
\end{align}
To find explicit solutions, we perform the change of variable (\ref{cv}),
then we get from (\ref{d1}), (\ref{d3}) and (\ref{d2}), the solutions for
the metric components
\begin{align}
g_{zz} & =\epsilon_{zz}\sigma^{2p}H^{2}\text{ },  \notag \\
g_{ww} & =\epsilon_{ww}\sigma^{2q}H^{2}\text{ },  \notag \\
g_{ab} & =h_{ac}\left( e^{2\lambda\log\sigma}\right) ^{c}{}_{b}H^{-\frac{2}{
D-3}}\text{ },  \label{metcomp}
\end{align}
where we have defined
\begin{equation}
\lambda _{\text{ \ }z}^{z}=p=\frac{1}{2}\theta _{\text{ }z}^{z}+1\text{ },
\text{ \ \ \ }\lambda _{\text{ \ }w}^{w}=q=\frac{1}{2}\theta _{\text{ }
w}^{w}+1\text{ },\text{ \ \ \ }\lambda _{\text{ \ }b}^{a}=\frac{1}{2}\theta
_{\text{ \ }b}^{a}-\frac{1}{D-3}\delta _{\text{ }b}^{a}\text{ }.  \label{kj}
\end{equation}
The constants $\epsilon _{zz}$ and $\epsilon _{ww}$ take the values $\pm 1$
and $h_{ac}$ are symmetric and satisfy (\ref{sym}).

Turning to (\ref{tr3}), we obtain in terms of the new variables
\begin{equation}
H\left( \frac{dH}{d\sigma}-\frac{\sigma}{H}\left( \frac{dH}{d\sigma}\right)
^{2}+\sigma\frac{d^{2}H}{d\sigma^{2}}\right) =-\frac{D-3}{2\left( D-2\right) 
}\varepsilon\epsilon_{0}\epsilon_{zz}\epsilon_{ww}\sigma^{-\left(
2p+2q-1\right) }P^{2}\text{ .}
\end{equation}
This admits the solution
\begin{equation}
H=1+c\sigma^{\xi}
\end{equation}
with
\begin{align}
\xi & =2\left( 1-p-q\right) \text{ }, \\
c& =-\frac{D-3}{2\left( D-2\right) \xi ^{2}}\varepsilon \epsilon
_{0}\epsilon _{zz}\epsilon _{ww}P^{2}\text{ }.  \label{xc}
\end{align}%
Using (\ref{kj}) then (\ref{w}) becomes 
\begin{equation}
\lambda _{\text{ \ }i}^{i}=p+q+\lambda _{\text{ \ }a}^{a}=1\text{ }.
\label{c}
\end{equation}
Finally (\ref{in}) gives the condition 
\begin{equation}
\frac{1}{2}\theta ^{i}{}_{j}\theta ^{j}{}_{i}+2\left( \frac{D-2}{D-3}\right)
\left( \theta _{\text{ }z}^{z}+\theta _{\text{ }w}^{w}+1\right) =0
\end{equation}
which in terms of $\lambda ^{i}{}_{j}$ simply implies
\begin{equation}
\lambda ^{i}{}_{j}\lambda ^{j}{}_{i}=p^{2}+q^{2}+\lambda ^{a}{}_{b}\lambda
^{b}{}_{a}=1\text{ }.  \label{cc}
\end{equation}
Collecting equations, our solutions are given by 
\begin{align}
ds^{2}& =H^{2}\left( \epsilon _{0}d\sigma ^{2}+\epsilon _{zz}\sigma ^{2p_{
\text{ }}}dz^{2}+\epsilon _{ww}\sigma ^{2q}dw^{2}\right) +h_{ac}\left(
e^{2\lambda \log \sigma }\right) ^{c}{}_{b}H^{-\frac{2}{D-3}}dx^{a}dx^{b}
\text{ },  \notag \\
H& =1-\frac{D-3}{8\left( D-2\right) \left( 1-p-q\right) ^{2}}\varepsilon
\epsilon _{0}\epsilon _{zz}\epsilon _{ww}P^{2}\sigma ^{2\left( 1-p-q\right) }
\text{ },  \notag \\
\text{\ }\mathcal{F}_{zw}& =P\text{ },  \label{gsss}
\end{align}%
with the conditions (\ref{c}) and (\ref{cc}).

For a given $\lambda _{\text{ }j}^{i}$, explicit solutions can be easily
obtained for both types of solutions (\ref{gs}) and (\ref{gsss}) for all
space-time dimensions. The possible choices of $\lambda _{\text{ }j}^{i}$
depend on their eigenvalues and their multiplicity.

\section{Examples}
In what follows, we will construct some explicit examples of the general
solutions (\ref{gs}) and (\ref{gsss}) in four and five space-time
dimensions. \ We start with the diagonal $\lambda _{\text{ }j}^{i}$ given by
\begin{equation}
\lambda ^{i}{}_{j}=\left( 
\begin{array}{ccc}
p & 0 & 0 \\ 
0 & a & 0 \\ 
0 & 0 & b%
\end{array}%
\right)   \label{aa}
\end{equation}%
where $p,$ $a$ and $b$ satisfying 
\begin{equation}
p+a+b=p^{2}+a^{2}+b^{2}=1\text{ },  \label{di}
\end{equation}%
then we obtain from (\ref{gs}) the four-dimensional metric which was first
considered in \cite{kt}, generalized to all signatures and is given by 
\begin{equation}
ds^{2}=H^{2}\left( \epsilon _{0}d\sigma ^{2}+\epsilon _{1}\sigma
^{2a}dx^{2}+\epsilon _{2}\sigma ^{2b}dy^{2}\right) +\frac{\epsilon
_{zz}\sigma ^{2p}}{H^{2}}dz^{2}
\end{equation}
with 
\begin{equation}
H=\left( 1+\varepsilon \epsilon _{zz}\frac{Q^{2}}{16p^{2}}\sigma
^{2p}\right) ^{2}\text{ },\text{ \ \ \ }\mathcal{F}^{\sigma z}=\frac{Q}{
\sigma H^{2}}\text{ }.  \label{dis}
\end{equation}
For the standard Einstein-Maxwell theory ($\varepsilon =1),$ choosing a
Lorentzian signature and relabelling the coordinates for the special case of 
$a=b=0$ and $p=1,$ one reproduces the well known Melvin solution 
\begin{align}
ds^{2}& =F^{2}(r)(-dt^{2}+dr^{2}+dy^{2})+{\frac{r^{2}}{F^{2}(r)}}d\varphi
^{2}\text{ },  \notag \\
\text{\ }\mathcal{F}^{r\varphi }& =\frac{Q}{rF^{2}},\quad F=1+\frac{Q^{2}}{16%
}r^{2}.
\end{align}%
with $\varphi $ being the azimuthal angle. Similarly one can obtain the so
called Melvin cosmology \cite{kt},
\begin{align}
ds^{2}& =F^{2}(t)(-dt^{2}+dx^{2}+dy^{2})+{\frac{t^{2}}{F(t)^{2}}}dz^{2}\text{
},\quad   \notag \\
\mathcal{F}^{tz}& =\frac{Q}{tF^{2}}\text{ },\quad F=1+\frac{Q^{2}}{16}t^{2}
\text{ }.
\end{align}
A new family of solutions can be obtained in four dimensions for a diagonal $
\lambda _{\text{ }j}^{i}$ where two of its eigenvalues are complex. Then $
\lambda _{\text{ }j}^{i}$ can be brought to the form 
\begin{equation}
\lambda _{\text{ }j}^{i}=\left( 
\begin{array}{ccc}
p & 0 & 0 \\ 
0 & a+ib & 0 \\ 
0 & 0 & a-ib
\end{array}
\right) \text{ },  \label{due}
\end{equation}
with the conditions 
\begin{equation}
p+2a=p^{2}+2a^{2}-2b^{2}=1\text{ }.
\end{equation}
Using the general solution (\ref{gs}), we obtain for this case a metric in a
complex form
\begin{equation}
ds^{2}=H^{2}\left( \epsilon _{0}d\sigma ^{2}+\epsilon _{1}\sigma
^{2a}e^{2ib\log \frac{\sigma }{\alpha }}\left( dx^{1}\right) ^{2}+\epsilon
_{2}\sigma ^{2a}e^{-2ib\log \frac{\sigma }{\alpha }}\left( dx^{2}\right)
^{2}\right) +\epsilon _{zz}\frac{\sigma ^{2p}}{H^{2}}dz^{2}\text{ }.
\label{cm}
\end{equation}
By performing the complex coordinate transformations,
\begin{equation}
x^{1}=\frac{1}{\sqrt{2}}\left( x-iy\right) \text{ },\ \ \ x^{2}=\frac{1}{
\sqrt{2}}\left( x+iy\right) \text{ },\ \ \ \ \ \ 
\end{equation}
with $\epsilon _{1}=\epsilon _{2},$ we can bring (\ref{cm}) to the real form
\begin{equation}
ds^{2}=\epsilon _{0}H^{2}d\sigma ^{2}+\epsilon _{1}H^{2}\sigma ^{2a}\left[
\cos \left( 2b\log \frac{\sigma }{\alpha }\right) \left(
dx^{2}-dy^{2}\right) +2\sin \left( 2b\log \frac{\sigma }{\alpha }\right) dxdy%
\right] +\epsilon _{zz}\frac{\sigma ^{2p}}{H^{2}}dz^{2}\text{ }.
\end{equation}

A further example can be obtained for $\lambda _{\text{ }j}^{i}$ with two
equal eigenvalues and can be brought to the form
\begin{equation}
\lambda _{\text{ }j}^{i}=\left( 
\begin{array}{ccc}
p & 0 & 0 \\ 
0 & a & 1 \\ 
0 & 0 & a
\end{array}
\right)  \label{3}
\end{equation}
with the conditions
\begin{equation}
p+2a=1\text{ },\text{ \ \ \ \ \ \ \ }2a^{2}+p^{2}=1\text{ }.
\end{equation}
In this case the metric solution takes the form
\begin{equation}
ds^{2}=H^{2}\left( \epsilon _{0}d\sigma ^{2}+2\epsilon _{1}\sigma
^{2a}dxdy+\epsilon _{1}\sigma ^{2a}\log \left( \frac{\sigma ^{2}}{\alpha }
\right) dy^{2}\right) +\epsilon _{zz}\frac{\sigma ^{2p}}{H^{2}}dz^{2}\text{ }
.
\end{equation}%
In all the above examples, the function $H$ and $\mathcal{F}^{\sigma z}$ are
given by (\ref{dis}).

As a first example in five dimensions, we consider the case where $\lambda _{
\text{ \ }j}^{i}$ is diagonal with real eigenvalues,
\begin{equation}
\lambda _{\text{ }j}^{i}=\left( 
\begin{array}{cccc}
p & 0 & 0 & 0 \\ 
0 & q & 0 & 0 \\ 
0 & 0 & a & 0 \\ 
0 & 0 & 0 & b%
\end{array}%
\right)  \label{kss}
\end{equation}
with the conditions
\begin{equation}
p+q+a+b=p^{2}+q^{2}+a^{2}+b^{2}=1\text{ }.  \label{ks}
\end{equation}
In this case one obtains the Kasner-like solutions 
\begin{equation}
ds^{2}=H^{2}\left( \epsilon _{0}d\sigma ^{2}+\epsilon _{1}\sigma
^{2q}dx^{2}+\epsilon _{2}\sigma ^{2a}dy^{2}+\epsilon _{3}\sigma
^{2b}dw^{2}\right) +\epsilon _{zz}\frac{\sigma ^{2p}}{H^{4}}dz^{2}\text{ }.
\end{equation}%
A new family of solutions can be also constructed for the choice 
\begin{equation}
\lambda _{\text{ }j}^{i}=\left( 
\begin{array}{cccc}
p & 0 & 0 & 0 \\ 
0 & q & 0 & 0 \\ 
0 & 0 & a+ib & 0 \\ 
0 & 0 & 0 & a-ib%
\end{array}
\right)  \label{al}
\end{equation}
with the conditions 
\begin{equation}
p+q+2a=p^{2}+q^{2}+2a^{2}-2b^{2}=1\text{ }.  \label{al1}
\end{equation}
The solutions can be brought to the following real form
\begin{align}
ds^{2}& =H^{2}\epsilon _{0}d\sigma ^{2}+\epsilon _{zz}\frac{\sigma ^{2p}}{
H^{4}}dz^{2}+\epsilon _{1}\sigma ^{2q}H^{2}d\omega ^{2}  \notag \\
& +2\epsilon _{2}\sigma ^{2a}H^{2}\sin \left( 2b\log \frac{\sigma }{\alpha }
\right) dxdy+\epsilon _{2}H^{2}\sigma ^{2a}\cos \left( 2b\log \frac{\sigma }{
\alpha }\right) \left( dx^{2}-dy^{2}\right) \text{ }.
\end{align}
Another class of solutions can be obtained for the choice
\begin{equation}
\lambda _{\text{ }j}^{i}=\left( 
\begin{array}{cccc}
p & 0 & 0 & 0 \\ 
0 & q & 0 & 0 \\ 
0 & 0 & a & 1 \\ 
0 & 0 & 0 & a%
\end{array}
\right)  \label{ca}
\end{equation}
with the conditions 
\begin{equation}
p+q+2a=1,\text{ \ \ \ \ \ \ \ }q^{2}+p^{2}+2a^{2}=1\text{ }.  \label{ca1}
\end{equation}
The metric in this case can take the form
\begin{equation}
ds^{2}=H^{2}\left( \epsilon _{0}d\sigma ^{2}+\epsilon _{1}\tau ^{2a}\log
\left( \frac{\sigma ^{2}}{\alpha }\right) dw^{2}+2\epsilon _{1}\sigma
^{2a}dydw+\epsilon _{2}\sigma ^{2q}dx^{2}\right) +\epsilon _{zz}\frac{\tau
^{2p}}{H^{4}}dz^{2}\text{ }.
\end{equation}%
In all the above examples we have 
\begin{equation}
H=\sqrt{1+\frac{\varepsilon \epsilon _{zz}}{12p^{2}}Q^{2}\sigma ^{2p}},\text{
\ \ \ }\mathcal{F}^{\sigma z}=\frac{Q}{\sigma H^{2}}\text{ }.
\end{equation}
Similarly one can construct solution for the choices 
\begin{equation}
\text{\ }\lambda _{\text{ }j}^{i}=\left( 
\begin{array}{cccc}
-\frac{1}{2} & 0 & 0 & 0 \\ 
0 & \frac{1}{2} & 1 & 0 \\ 
0 & 0 & \frac{1}{2} & 1 \\ 
0 & 0 & 0 & \frac{1}{2}%
\end{array}%
\right) \text{ },\text{ \ \ }\lambda _{\text{ }j}^{i}=\left( 
\begin{array}{cccc}
1 & 0 & 0 & 0 \\ 
0 & 0 & 1 & 0 \\ 
0 & 0 & 0 & 1 \\ 
0 & 0 & 0 & 0
\end{array}
\right) \text{ },
\end{equation}
which is left as an exercise to the reader.

For the second class of solutions with \ $\mathcal{F}_{zw}=P$ and represented by (\ref{gsss}), we first consider examples in four dimensions.
Here we have only one possibility for $\lambda ^{i}{}_{j}$ given by (\ref{aa}) with the conditions (\ref{di}). The solution obtained can be represented
by 
\begin{align}
ds^{2}& =H^{2}\left( \epsilon _{0}d\sigma ^{2}+\epsilon _{zz}\sigma
^{2p}dz^{2}+\epsilon _{ww}\tau ^{2a}dw^{2}\right) +\epsilon _{3}\frac{\tau
^{2b}}{H}dx^{2},  \notag \\
H& =1-\frac{1}{16b^{2}}\varepsilon \epsilon _{0}\epsilon _{zz}\epsilon
_{ww}P^{2}\sigma ^{2b}\text{ }.
\end{align}
In five dimensions, as a first example we consider the case where $\lambda _{
\text{ \ }j}^{k}$ is given by (\ref{kss}) with the conditions (\ref{ks}). \
This produces the solution 
\begin{equation}
ds^{2}=H^{2}\left( \epsilon _{0}d\sigma ^{2}+\epsilon _{zz}\sigma
^{2p}dz^{2}+\epsilon _{ww}\sigma ^{2q}dw^{2}\right) +H^{-1}\left( \epsilon
_{3}\sigma ^{2a}dx^{2}+\epsilon _{4}\sigma ^{2b}dy^{2}\right)
\end{equation}
with 
\begin{equation}
H=1-\frac{1}{12\left( a+b\right) ^{2}}\epsilon _{0}\epsilon \epsilon
_{zz}\epsilon _{ww}P^{2}\sigma ^{2\left( a+b\right) }\text{ }.
\end{equation}%
Next we consider the case for $\lambda _{\text{ }j}^{i}$ given by (\ref{al})
with the conditions (\ref{al1}), then we obtain the solution
\begin{align}
ds^{2}& =H^{2}\left( \epsilon _{0}d\sigma ^{2}+\epsilon _{zz}\sigma
^{2p}dz^{2}+\epsilon _{ww}\sigma ^{2q}dw^{2}\right)   \notag \\
& +2\epsilon _{1}\tau ^{2a}H^{-1}\sin \left( 2b\log \frac{\sigma }{\alpha }
\right) dydx+\epsilon _{1}H^{-1}\tau ^{2a}\cos \left( 2b\log \frac{\sigma }{
\alpha }\right) \left( dy^{2}-dx^{2}\right) \text{ }.
\end{align}
The case for $\lambda _{\text{ }j}^{i}$ given by (\ref{ca}) with the
conditions (\ref{ca1}) produces the solution
\begin{align}
ds^{2}& =H^{2}\left( \epsilon _{0}d\tau ^{2}+\epsilon _{1}\tau
^{2p}dz^{2}+\epsilon _{2}\tau ^{2q}dw^{2}\right) +H^{-1}\epsilon _{3}\tau
^{2a}\left( 2dxdy+\log \frac{\sigma ^{2}}{\alpha }dy^{2}\right) \text{ }, 
\notag \\
H& =1-\frac{1}{48a^{2}}\varepsilon \epsilon _{0}\epsilon _{zz}\epsilon
_{ww}P^{2}\sigma ^{4a}\text{ }.
\end{align}
The constants $\epsilon _{0},$ $\epsilon _{zz},$ $\epsilon _{ww}$ and $
\epsilon _{i}$ take the values $\pm 1$ for all the solutions and $\alpha $
is an arbitrary constant.

In summary, we have obtained new solutions depending on one variable for the
theories of $D$-dimensional Einstein-Maxwell theory in arbitrary space-time
signature. The general form of the two classes of solutions are given by (\ref{gs}) and (\ref{gsss}).

Apart from the diagonal Kasner-like solution, a common feature among both
types of the new solutions is that the transverse metric $g_{ij}$ can not be
made Euclidean for all choices of the coordinates. As such, domain wall
solutions with Lorentzian signature can always be constructed but the
cosmological solutions are necessarily non-Lorentzian. This implies that the
correspondence between domain walls and cosmological solutions discussed in 
\cite{kt, s1}, with our new solutions, is extended to solutions
corresponding to different space-time signatures. This is in contrast to the
Kasner-like solutions of \cite{kt, s1} where Lorentzian domain walls can be
related to Lorentzian cosmological solutions.

Our results can be generalized to find new solutions to supergravity
theories in various space-time dimensions and space-time signatures. Also in
light of the AdS/CFT\ correspondence, it is important to find and analyze
solutions to Einstein-Maxwell theory with a cosmological constant and to
general gauged supergravity models. Work in these directions is in progress
and we hope to report on it in the near future.

\bigskip

{\flushleft{\textbf{Acknowledgements:}}} I am grateful to J. B. Gutowski for
many helpful discussions.\vfill\eject

\end{document}